\documentclass[preprint,%reprint,
superscriptaddress,aps,prl]{revtex4-1}

\usepackage{graphicx}% Include figure files
\usepackage{dcolumn,booktabs}% Align table columns on decimal point
\newcolumntype{d}[1]{D{.}{.}{#1}}
\newcommand\mc[1]{\multicolumn{1}{c}{#1}}

\begin{document}

\title{Revisiting the Photon-Drag Effect in Metal Films}

\author{Jared H. Strait}
\email{jared.strait@nist.gov}
\affiliation{Physical Measurement Laboratory, National Institute of Standards and Technology, Gaithersburg, MD 20899}
\author{Glenn Holland}
\affiliation{Physical Measurement Laboratory, National Institute of Standards and Technology, Gaithersburg, MD 20899}
\author{Wenqi Zhu}
\affiliation{Physical Measurement Laboratory, National Institute of Standards and Technology, Gaithersburg, MD 20899}
\affiliation{Maryland Nanocenter, University of Maryland, College Park, MD 20742}
\author{Cheng Zhang}
\affiliation{Physical Measurement Laboratory, National Institute of Standards and Technology, Gaithersburg, MD 20899}
\affiliation{Maryland Nanocenter, University of Maryland, College Park, MD 20742}
\author{Bojan R. Ilic}
\affiliation{Physical Measurement Laboratory, National Institute of Standards and Technology, Gaithersburg, MD 20899}
\author{Amit Agrawal}
\affiliation{Physical Measurement Laboratory, National Institute of Standards and Technology, Gaithersburg, MD 20899}
\affiliation{Maryland Nanocenter, University of Maryland, College Park, MD 20742}
\author{Domenico Pacifici}
\affiliation{Physical Measurement Laboratory, National Institute of Standards and Technology, Gaithersburg, MD 20899}
\affiliation{School of Engineering and Department of Physics, Brown University, Providence, RI 02906}
\author{Henri J. Lezec}
\affiliation{Physical Measurement Laboratory, National Institute of Standards and Technology, Gaithersburg, MD 20899}

\date{\today}

\begin{abstract}
The photon-drag effect, the rectified current in a medium induced by conservation of momentum of absorbed or redirected light, is a unique probe of the detailed mechanisms underlying radiation pressure.
We revisit this effect in gold, a canonical Drude metal.
We discover that the signal for $p$-polarized illumination in ambient air is affected in both sign and magnitude by adsorbed molecules, opening previous measurements for reinterpretation.
Further, we show that the intrinsic sign of the photon-drag effect is contrary to the prevailing intuitive model of direct momentum transfer to free electrons.
\end{abstract}

\maketitle

Transfer of momentum is a fundamental aspect of light-metal interaction thought to be well understood within the framework of classical electrodynamics.
For instance, the Lorentz force density predicts net momentum balance between light waves and a highly-conductive material \cite{planck_theory_1914}.
The myriad calibrated demonstrations of radiation pressure, dating to the early twentieth century \cite{nichols_preliminary_1901}, confirm this momentum balance but do not explicitly probe
the microscopic processes responsible for momentum transfer.
Since the infrared conductivity in noble metals such as Au and Ag is dominated by Drude-like free carriers \cite{olmon_optical_2012}, it would stand to reason that momentum absorbed from infrared light transfers initially to the free electrons.
The resulting rectified current flow, induced by the momentum of an electromagnetic wave, is known as the photon-drag effect \cite{wegener_extreme_2005, loudon_radiation_2005, goff_hydrodynamic_1997, goff_theory_2000}.
Let $p_\|$ be the in-plane component of the absorbed light momentum.  
The in-plane voltage across the illuminated metal area is then $V_\|=\xi\ \! p_\|$, where the voltage transduction factor $\xi$ depends on the microscopic processes of momentum exchange \cite{goff_hydrodynamic_1997, goff_theory_2000, gurevich_photomagnetism_1992}.
Thus, photon drag is a unique probe of light-metal interaction, sensitive not only to net momentum balance but also to underlying momentum exchange mechanisms.

Recent interest in the metallic photon-drag effect has focused on the role of plasmons, motivated by the ability to engineer the light-metal interaction and to investigate a novel platform for electrical detection of plasmons in nanophotonic systems with multi-terahertz bandwidth \cite{durach_giant_2009, bai_manipulating_2015, durach_spin_2017}.
Photocurrents related to plasmonic interactions are observed for metal films in the Kretschmann-Raether (KR) configuration \cite{lupi_electrical_2014, vengurlekar_surface_2005, noginova_light--current_2011, durach_nature_2016} and with nanostructured or roughened surfaces \cite{noginova_plasmon_2013, hatano_transverse_2009, akbari_photo-induced_2015, hatano_optical_2008, kurosawa_optical_2012, kurosawa_surface_2012, noginova_plasmonic_2016, proscia_control_2016, akbari_polarization_2017, akbari_generation_2018}.
But the direction of the measured current has been reported to vary with the excitation conditions in a manner inconsistent with the expectation of a unique direction of momentum transfer \cite{vengurlekar_surface_2005, hatano_transverse_2009, noginova_light--current_2011, noginova_plasmon_2013, akbari_photo-induced_2015, durach_nature_2016}.
For nominally-smooth metal films, reported photon-drag currents generated at the KR condition \cite{lupi_electrical_2014, vengurlekar_surface_2005, noginova_light--current_2011, durach_nature_2016} and the air-metal interface using $p$-polarized light \cite{noginova_plasmon_2013} both imply electron flow parallel to (in the direction of) $p_\|$, as expected for direct momentum transfer to free electrons.
But reported photocurrents generated at the glass-metal interface away from the KR angle \cite{vengurlekar_surface_2005, noginova_light--current_2011, noginova_plasmon_2013, durach_nature_2016}, and for $s$ waves at the air-metal interface \cite{hatano_transverse_2009}, have a sign implying electron flow \emph{antiparallel} to $p_\|$, opposite to what momentum conservation intuitively implies.
Thus the sign of the photon-drag effect in metals, a fundamental characteristic of light-metal interaction, beckons further investigation.

In this Letter, we experimentally revisit the photon-drag effect in thin metal films to investigate the discrepancies found in the literature.  
By employing smooth gold films in a vacuum environment, we return to an intentionally simple configuration devoid of plasmonic nanostructures that might obscure the origins of reported signals.
We find that the photocurrent measured in vacuum is proportional to $p_\|$, consistent with the photon-drag effect, and the currents excited with both light polarizations and at both surfaces of the film all have the same sign.  
But we show that when the experiment is repeated in an ambient air environment, the $p$-wave photocurrent is dominated by molecular surface adsorbates, revealing the sign inconsistencies discussed above.  
Furthermore, this extrinsic, environment-dependent electron flow is deceptively parallel to $p_\|$, whereas the intrinsic photon-drag electron flow is counterintuitively antiparallel.  
These conclusions (1) require a reassessment of the previous metallic photon-drag experiments, most of which were performed in air \cite{lupi_electrical_2014, vengurlekar_surface_2005, noginova_light--current_2011, durach_nature_2016, noginova_plasmon_2013, hatano_optical_2008, hatano_transverse_2009, kurosawa_optical_2012, kurosawa_surface_2012, akbari_photo-induced_2015, noginova_plasmonic_2016, proscia_control_2016, akbari_polarization_2017, akbari_generation_2018}, (2) suggest an optoelectronic paradigm for molecular sensing on metal surfaces, and (3) necessitate a reimagination of the physical processes responsible for transduction of electrical currents following momentum transfer in light-metal interactions.

\begin{figure}[tb]
  \centering
  \includegraphics[width=3.25 in]{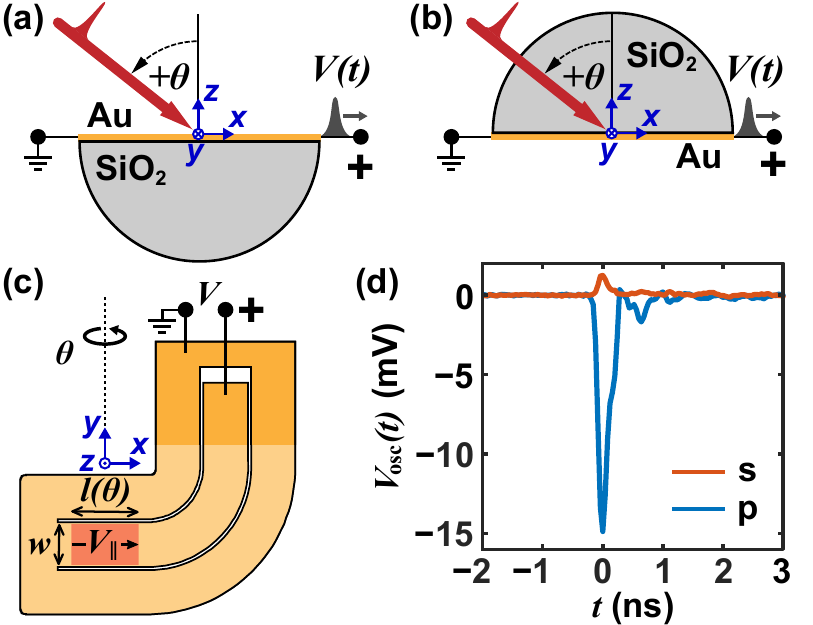}
\caption{(a-b) Schematic illustration of the experiment for illumination at the (a) free space-metal or (b) glass-metal interface.
``+" refers to the amplifier input.
Plane of metal: $x$-$y$; plane of incidence: $x$-$z$; $\theta$ rotation axis: $y$.
(c) Device layout showing Au waveguide: light yellow; ohmic contact area: dark yellow; illuminated area: red.
(d) Representative measured waveforms for $s$- and $p$-polarized illumination at the free space-metal interface with $\theta=+45^\circ$ in ambient air.}
\label{CartoonTimeEnergy}
\end{figure}

Our examination of the photon-drag effect began with measuring the photovoltage across a gold film as a function of laser polarization (linear, $s$ or $p$), angle of incidence $\theta$, ambient air pressure, and incident medium.
The gold, chosen for the simple monovalent $s$-band character of its Fermi surface \cite{ashcroft_solid_1976} and Drude-like conductivity in the infrared range \cite{olmon_optical_2012}, was deposited with no adhesion layer on a fused-silica substrate using ion-beam sputtering.
Our film had thickness $d\approx35$ nm, with $<\! 1$ nm surface roughness.
To access large in-plane wavevectors at the glass-metal interface, we optically contacted a hemicylindrical prism to the reverse of the substrate, as pictured in Fig.~\ref{CartoonTimeEnergy}(a-b).
The Au film was patterned into a shorted coplanar waveguide (Fig.~\ref{CartoonTimeEnergy}(c)), including an elbow curve to provide a compact form factor, with track width $w=4$ mm and gap of 150 $\mu$m to provide a waveguide impedance of $\sim$50 $\Omega$.
A 150-nm thick Au pad provides ohmic contact to the photon-drag voltage-detection circuit, consisting of a 10 GHz bandwidth amplifier (voltage amplification factor $\Gamma=18$) and an 8 GHz bandwidth oscilloscope.
The device was mounted into a vacuum chamber which was either vented with ambient air (293 K, relative humidity 15 \%-20 \%) or evacuated to $\approx\!10^{-3}$ Pa for several hours under illumination of an ultraviolet C (UVC) mercury discharge lamp for water desorption.  
An optical pulse train, from a 5 mJ, 1 kHz, 800 nm wavelength Ti:sapphire chirped-pulse amplifier, with rectangular beam profile of width $w_{\rm p}=w$ and length $l_{\rm p}=1$ mm, illuminated an area of the Au film of constant width $w$ and variable length $l(\theta)=l_p/\cos\theta$. 
The pulses were left uncompressed with duration  $t_{\rm p}\approx 20$ ps to suppress multiphoton photoemission occurring at high fields \cite{lompre_new_1978}.
All signal amplitudes were linearly proportional to optical pulse energy, consistent with a momentum-driven effect.

An optical pulse generates an in-plane current pulse $I_\|(t)$ in the Au film, either parallel or antiparallel to $p_\|$, and a corresponding potential drop across the illuminated area $V_\|(t)$.
The pulse propagates down the waveguide and traverses lossy contacts to the amplifier, yielding an attenuated and broadened voltage pulse at the amplifier input $V(t)$ (Fig.~\ref{CartoonTimeEnergy}(a-c)).
The waveguide geometry is equivalent to the single-wire geometries used in previous reports \cite{vengurlekar_surface_2005, hatano_transverse_2009, noginova_plasmon_2013} for the sign of the transduced signal, and we electrically verified that the detection circuit does not invert the sign of a pulse \cite{noauthor_see_nodate}.
Representative amplified waveforms measured by the oscilloscope $V_{\rm osc}(t)$ appear as unipolar pulses of width $t_V\approx300$ ps (Fig.~\ref{CartoonTimeEnergy}(d)).  
We report the peak amplitude of these waveforms, divided by $\Gamma$ to find the peak voltage at the output of the waveguide, $V_j^i(\theta)$.
Here $i=s,p$ indexes the optical polarization, and $j=f,g$ indexes the free space-metal and glass-metal illumination surfaces, respectively.

\begin{figure*}[tb]
  \centering
  \includegraphics[width=6.48in]{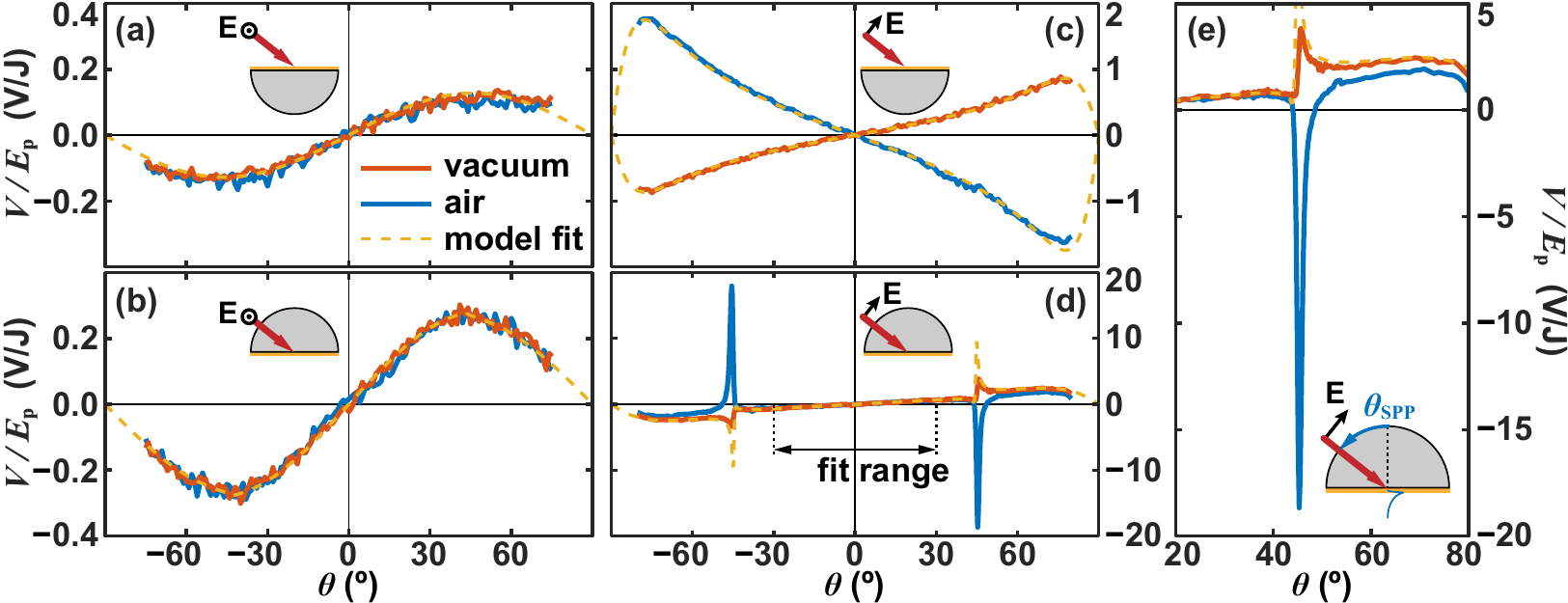}
\caption{Dependence on angle of incidence for $s$-polarization (a,b) and $p$-polarization (c-e). 
(a) $V_f^s(\theta)/E_{\rm p}$, (b) $V_g^s(\theta)/E_{\rm p}$, (c) $V_f^p(\theta)/E_{\rm p}$, and (d-e) $V_g^p(\theta)/E_{\rm p}$ in ambient air and vacuum, along with $\xi_j^i p_\|(\theta)/E_{\rm p}$ fits.
(e) Enlarged view of experimental data in (d) around +$\theta_{\rm SPP}$.
%Angular resolution is $\approx\!1.4^\circ$.
}
\label{AngleCombine}
\end{figure*}

Plots for $V_j^i(\theta)$, measured in vacuum and air environments and normalized to the optical pulse energy $E_{\rm p}$, appear in Fig.~\ref{AngleCombine}.
We model the data curves with $\xi_j^i p_\|(\theta)/E_{\rm p}$, where $p_\|(\theta)$ is the absorbed in-plane component of the laser pulse momentum in vacuum, using the voltage transduction factor at waveguide output $\xi_j^i$ as the lone fitting parameter (collected in Table~\ref{xiTable}).
Since strong deviations from the model exist for $V_g^p(\theta)$ around $\theta=\pm45^\circ$, we restrict the fit of $\xi_g^p$ to the range $-30^\circ\le\theta\le30^\circ$ for this case.
We compute the optical power absorption coefficients $A_j^i(\theta)$ for the Au film with the Fresnel formulas \cite{kats_optical_2016},
using the Au refractive index $n_{\rm Au}=0.31 + 4.88 i$ at $\lambda=800$ nm measured by ellipsometry. 
Then $p_\|(\theta)=(E_{\rm p}/c)A(\theta)\sin(\theta)$, where $c$ is the speed of light in vacuum and indices are omitted for clarity.
The quality of these fits ($R^2>0.96$) indicates that the measured $V_j^i(\theta)$ are primarily proportional to $p_\|(\theta)$ as expected for the photon-drag effect.  

\begin{table}[tbp]
\begin{ruledtabular}
\begin{tabular}{ l  *{4}{d{3.3}} }
    		& 	\mc{$\xi_f^s$}	&	\mc{$\xi_g^s$}	&	\mc{$\xi_f^p$}	&	\mc{$\xi_g^p$}	\\	\hline
	Vacuum	&	1.14		&	1.74		&	1.95		&	4.21		\\
    Air		&	1.12		&	1.72		&	-3.95		&	3.91		\\
\end{tabular}
\end{ruledtabular}
\caption{Fitted voltage transduction factors $\xi_j^i$, measured at the amplifier input (uncorrected for propagation loss), in units of GV/(N$\:$s) for photovoltages measured in vacuum and ambient air environments.
Fit 95~\% confidence interval, computed as twice the standard error, is approximately $\pm0.08$ GV/(N$\:$s) for $\xi_g^p$ and $\pm0.03$ GV/(N$\:$s) for $\xi_j^s$ and $\xi_f^p$.
But systematic errors limit the total confidence interval to approximately $\pm 5$~\% $\xi_j^i$.
}
\label{xiTable}
\end{table}

\begin{figure}[tb]
  \centering
  \includegraphics[width=3.25in]{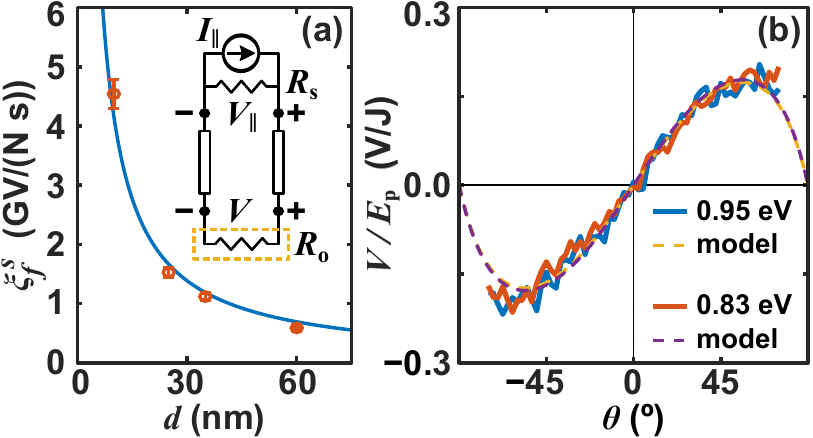}
\caption{(a) $\xi_f^s$ versus thickness of the Au film, along with a fit proportional to $1/d$ ($R^2=0.979$). 
Inset: Diagram of measurement circuit.
Dashed box represents amplifier and oscilloscope.
(b) $V_g^s(\theta)/E_{\rm p}$ for illumination with photon energies of 0.95 eV and 0.83 eV, along with $\xi_g^s p_\| (\theta)/E_{\rm p}$ fits.
Extracted $\xi(0.95$~eV$)=(4.6\pm 0.23)$ GV/(N$\:$s) and $\xi(0.83$~eV$)=(5.2\pm 0.26)$ GV/(N$\:$s), where the uncertainty includes the 95 \% confidence interval and systematic errors.}
\label{ThicknessColor}
\end{figure}

Positive angle of incidence $\theta$ is defined such that the in-plane incident light momentum points toward the input of the non-inverting amplifier.
Thus, the prevailing intuition of photon drag voltage transduction would imply a negative $V_j^i(\theta)$ for positive $\theta$, since it is expected that electrons would be ``pushed'' toward the amplifier input \cite{goff_hydrodynamic_1997, goff_theory_2000, gurevich_photomagnetism_1992, vengurlekar_surface_2005, noginova_plasmon_2013, durach_nature_2016}.
Under this assumption of direct momentum transfer to free electrons in the relaxation time approximation, the light-generated current through the Au cross-section is \cite{goff_theory_2000}
\begin{equation}
I_{\|,\rm expect}=w\,\sigma\left(\frac{\langle S\rangle}{N\,e\,c}\right)A(\theta)\sin(\theta)\cos(\theta),
\label{eq1}
\end{equation}
Here, $\langle S\rangle$ is the incident time-averaged Poynting flux, $\sigma$ is the Au conductivity, $N=5.9\times 10^{22}$ cm$^{-3}$ \cite{ashcroft_solid_1976} is the free-electron density, and $e<0$ is the electron charge.
This current is driven into two paths (see Fig.~\ref{ThicknessColor}(a) inset): (a) the detection circuit, including the waveguide, contacts, amplifier (input resistance $R_{\rm o}$), and oscilloscope and (b) a parallel shunt resistance $R_{\rm s}=l_{\rm p}/(\sigma w d\cos\theta)$ provided by the illuminated metal film itself.
Since $R_{\rm s}$ ($<$1 $\Omega$ for $\theta<80^\circ$) is low compared to $R_{\rm o}$ (50 $\Omega$), the potential presented to the circuit outside the illuminated area simplifies to $V_{\|,\rm expect}=I_{\|,\rm expect}\,l_{\rm p}/(\sigma wd\cos\theta)$.
Combining with Eq.~\ref{eq1} and using $\langle S\rangle=E_{\rm p}/(l_{\rm p} w t_{\rm p})$, we find
\begin{equation}
V_{\|,\rm expect} = \frac{p_\|}{w d t_{\rm p} N e} = \xi_{\rm expect}\,p_\|.
\label{eq2}
\end{equation}
Thus, the voltage transduction factor expected at the illuminated area, under the assumption of direct momentum transfer to free electrons, is $\xi_{\rm expect}=(w d t_{\rm p} N e)^{-1}=-38$ GV/(N$\:$s) for this device (yielding an expected transduction factor at waveguide output of $-2.5$ GV/(N$\:$s), assuming only lossless broadening $t_{\rm p}\to t_V$).

The measured transduction factors $\xi_j^i$ consistent with $\xi_{\rm expect}$ would then be negative and inversely proportional to thickness.
Indeed $\xi_f^s\propto d^{-1}$ as shown in Fig.~\ref{ThicknessColor}(a), verifying this embedded-current-source behavior of the effect.
But as seen in Table~\ref{xiTable}, only $\xi_f^p$ measured in air has the expected negative sign.
For all other configurations---seven out of eight, including $\xi_f^p$ measured in vacuum---the extracted $\xi$ are positive,
implying that the fundamental response of the metal to momentum transfer yields a net free-electron flow antiparallel to $p_\|$.
$P$-wave illumination of the free space-metal interface is the only configuration for which $\xi$ depends significantly, in both sign and magnitude, on the environment.  
We therefore argue that in ambient air, molecules at or near the Au surface activate an effect which dominates the $p$-wave-generated current for free-space illumination, resulting in $\xi_f^p<0$.
Since $\textrm{sgn}(\xi_f^p)=\textrm{sgn}(\xi_{\rm expect})$, it would be tempting to view this case as representative of the fundamental photon-drag transduction process, but our measurements in vacuum suggest otherwise.
This result undermines the interpretation of many reported photon drag measurements on metal films in air \cite{lupi_electrical_2014, vengurlekar_surface_2005, noginova_light--current_2011, durach_nature_2016, noginova_plasmon_2013, hatano_optical_2008, hatano_transverse_2009, kurosawa_optical_2012, kurosawa_surface_2012, akbari_photo-induced_2015, noginova_plasmonic_2016, proscia_control_2016, akbari_polarization_2017, akbari_generation_2018}, which implicitly treat the metal surface as pristine and the measured signals, therefore, as wholly intrinsic.

A surface plasmon polariton (SPP) can be excited at the free space-metal interface under the KR coupling condition \cite{raether_surface_1988}.
Evidenced by peaks in the calculated absorption at $\theta_{\rm SPP}=\pm45.0^\circ$, this condition also yields pronounced peaks in photovoltage $V_g^p(\theta\approx\pm\theta_{\rm SPP})$ which flip sign as the environment is changed between air and vacuum.
Thus, the sign of the enhanced photon-drag signal at the SPP coupling condition measured in vacuum is inconsistent with the intuition \cite{vengurlekar_surface_2005, noginova_light--current_2011, durach_nature_2016} for direct momentum exchange between the propagating SPP mode and free electrons.

The signals measured in ambient air, shown in Fig.~\ref{AngleCombine}, reproduce the signs (and sign discrepancies) of analogous experiments in the literature \cite{vengurlekar_surface_2005, noginova_light--current_2011, noginova_plasmon_2013, durach_nature_2016, hatano_transverse_2009}.
This agreement with independent experimental groups corroborates our measured signal signs and shows that our findings are unlikely to be the result of a quirk of measurement configuration.

Unlike in the air environment, all photovoltages measured for $\theta>0$ in vacuum are positive.
This finding eliminates sign discrepancies, but demonstrates that the intrinsic signal has a sign in conflict with the prevailing intuition for photon drag.  
We are left to conclude that the observed electron flow antiparallel to $p_\|$ results from either (a) direct momentum transfer from photons to electrons, along with a counterintuitive current transduction, or (b) an altogether different optical rectification process having the same $\theta$ dependence.
Although alternative second-order rectification channels for $p$-polarization are known, including surface optical rectification \cite{kadlec_study_2005, hubner_theory_1994} and surface-scattering photogalvanism \cite{gurevich_photomagnetism_1993, gurevich_photomagnetism_2000, mikheev_interplay_2018}, these processes could not explain the case of $s$-polarization.
Also, the positive sign of $\xi$ is reminiscent of the resonant photon-drag effect observed in semiconductors \cite{gibson_study_1980, wieck_observation_1990}, where counterintuitive signals were attributed to the details of interband transitions \cite{gibson_theoretical_1975, luryi_photon-drag_1987, grinberg_theory_1988, shalaev_light-induced_1996}.
But as shown in Fig.~\ref{ThicknessColor}(b), $\xi$ retains a positive sign even for excitation photon energies well below 1.6 eV, where interband transitions in Au are negligible and conductivity is dominated by intraband scattering \cite{olmon_optical_2012}.
We therefore rule out contributions from this resonant effect.
In the absence of other known rectification processes, conclusion (b) is unlikely for $s$-wave illumination.
Further evidence to endorse conclusion (a) lies in the comparison between $\xi_g^s$ and $\xi_f^s$.
If the photovoltage were simply proportional to absorbed optical energy, then the expectation would be $\xi_g^s=\xi_f^s$.
But instead we measure $\xi_g^s/\xi_f^s=1.53\pm0.11$, statistically indistinguishable from the refractive index of the fused-silica prism $n=1.453$.
This suggests that the $s$-wave photovoltage is proportional to the Minkowski photon momentum in the fused-silica \cite{barnett_resolution_2010, partanen_photon_2017}, $n\hbar\omega/c$, and can be interpreted as a photon-drag signal representing a direct measurement of the electromagnetic momentum incident upon the metal.
For the $p$-wave photovoltage, photon drag remains only one of several possibilities.

We now highlight three remaining questions concerning the results in Table~\ref{xiTable}: 
(i) What is the mechanism responsible for the environmental dependence of $V_f^p(\theta)$ and $V_g^p(\theta\approx\theta_{\rm SPP})$?
(ii) Why is the environmentally-insensitive transduction factor $\xi_g^p$ anomalously larger than $\xi_g^s$, with $\xi_g^p/\xi_g^s\approx2.3$ %\pm0.2$
, contradicting the expectation that different linear polarization states of light carry the same momentum?
(iii) Why does the photon-drag signal have a positive sign for $\theta>0$, implying electron flow antiparallel to $p_\|$?

\begin{figure}[tb]
  \centering
  \includegraphics[width=3.25in]{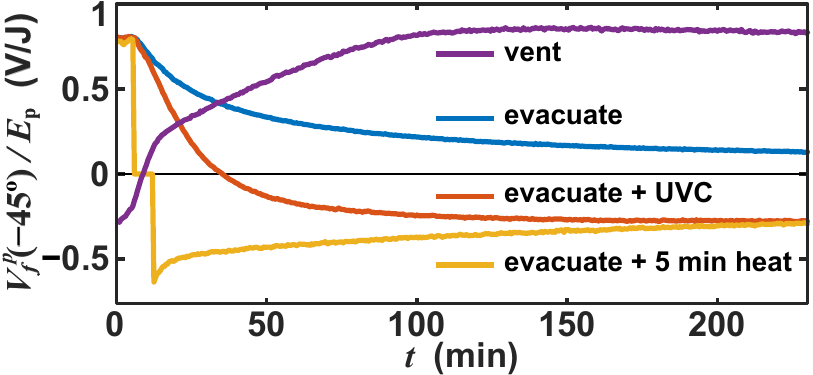}
\caption{
$V_f^p(-45^\circ)/E_{\rm p}$ versus time during environmental procedures: evacuation alone, evacuation with UVC illumination, evacuation with Joule heating (5 min during which the device was disconnected), and venting with ambient air.
}
\label{Timeseries}
\end{figure}

To investigate question (i), we collected a time series of the photovoltage signals while evacuating and subsequently venting the vacuum chamber.
During evacuation procedures, the turbomolecular pump was activated at time $t=0$ (at atmospheric pressure), and the UVC illumination or Joule heating began at $t=6$ minutes. %, once the pressure had decreased below $10^{-2}$ Pa.
Under evacuation, $V_f^p(-45^\circ)$ decreases toward zero and relaxes to a new steady state over hundreds of minutes (see Fig.~\ref{Timeseries}).
Including UVC illumination during evacuation increases the rate of change of $V_f^p(-45^\circ)$, which ultimately reaches a steady state with negative sign.
Joule heating of the Au film by dissipating 1 W of electrical power over 5 minutes in vacuum (raising the temperature to $\approx\!400$ K) causes a fast sign change of $V_f^p(-45^\circ)$ and even a slow relaxation in the positive direction as adsorbates return to the surface in the $\approx\!10^{-3}$ Pa vacuum.
When reexposed to the air environment during the vent procedure (at $t=0$ the environment reaches atmospheric pressure), $V_f^p(-45^\circ)$ recovers its positive sign, again over hundreds of minutes.
These slow dynamics are consistent with chemical adsorption and desorption of water from the Au surface \cite{wells_adsorption_1972}, a hypothesis further corroborated by the effects of UVC illumination and Joule heating.
The $p$-waves photovoltage is clearly a sensitive detector of surface adsorbates, likely water, at the submonolayer level \cite{demirdjian_water_2018}. 
Although the microscopic mechanism of this adsorbate-sensitive rectified current remains unclear, it might be regarded as the dc facet of an adsorbate-sensitive sum-frequency generation process at the metal surface \cite{brown_effect_1969, shen_surface_1989}.

Explanations of questions (i), environmental sensitivity, and (ii), anomalous ratio of $\xi_g^p/\xi_g^s$, might both lie in the surface-dependent effects allowed by symmetry for $p$-waves \cite{kadlec_study_2005, hubner_theory_1994, gurevich_photomagnetism_1993, gurevich_photomagnetism_2000, mikheev_interplay_2018}.
Alternatively, potentially differing optical force density distributions between the polarizations \cite{mansuripur_electromagnetic-force_2013} might lead to corresponding differences in voltage transduction.
Further inquiry will be needed to resolve these questions.
Lastly, the counterintuitive photon-drag current, 
question (iii), is revealed by our environmentally-pristine measurements and indicates the unique power of this technique to probe light-matter interaction.
Voltage measurements are sensitive to the currents generated during the intermediate processes of momentum transfer---the electrical echoes of radiation pressure.

Classical electromagnetic theory is united with the momentum-balance principle in predicting that the in-plane component of radiation pressure on the metal will be parallel to $p_\|$, not antiparallel \cite{gurevich_photomagnetism_1992,goff_hydrodynamic_1997,zangwill_modern_2012}.
But overall momentum conservation is not at question.  
Rather, we argue that the \emph{transduction} of a rectified current resulting from radiation pressure may be more subtle than originally thought.  
By what microscopic mechanisms does metal receive momentum from an electromagnetic wave?
If, as is generally assumed, the screening electrons are initially the sole objects of radiation pressure in a metal, then the photon-drag current would be consistent with momentum balance.
But the measured $\xi$ show that the intrinsic photon-drag current is antiparallel to $p_\|$ for gold, evidencing a previously unrecognized physical process at the heart of momentum transfer between light and matter.  
In this regard, the photon-drag effect is in a unique position to experimentally clarify the physical picture of this ubiquitous interaction and usher a new microscopic theory of radiation pressure.


\begin{thebibliography}{49}%
\makeatletter
\providecommand \@ifxundefined [1]{%
 \@ifx{#1\undefined}
}%
\providecommand \@ifnum [1]{%
 \ifnum #1\expandafter \@firstoftwo
 \else \expandafter \@secondoftwo
 \fi
}%
\providecommand \@ifx [1]{%
 \ifx #1\expandafter \@firstoftwo
 \else \expandafter \@secondoftwo
 \fi
}%
\providecommand \natexlab [1]{#1}%
\providecommand \enquote  [1]{``#1''}%
\providecommand \bibnamefont  [1]{#1}%
\providecommand \bibfnamefont [1]{#1}%
\providecommand \citenamefont [1]{#1}%
\providecommand \href@noop [0]{\@secondoftwo}%
\providecommand \href [0]{\begingroup \@sanitize@url \@href}%
\providecommand \@href[1]{\@@startlink{#1}\@@href}%
\providecommand \@@href[1]{\endgroup#1\@@endlink}%
\providecommand \@sanitize@url [0]{\catcode `\\12\catcode `\$12\catcode
  `\&12\catcode `\#12\catcode `\^12\catcode `\_12\catcode `\%12\relax}%
\providecommand \@@startlink[1]{}%
\providecommand \@@endlink[0]{}%
\providecommand \url  [0]{\begingroup\@sanitize@url \@url }%
\providecommand \@url [1]{\endgroup\@href {#1}{\urlprefix }}%
\providecommand \urlprefix  [0]{URL }%
\providecommand \Eprint [0]{\href }%
\providecommand \doibase [0]{http://dx.doi.org/}%
\providecommand \selectlanguage [0]{\@gobble}%
\providecommand \bibinfo  [0]{\@secondoftwo}%
\providecommand \bibfield  [0]{\@secondoftwo}%
\providecommand \translation [1]{[#1]}%
\providecommand \BibitemOpen [0]{}%
\providecommand \bibitemStop [0]{}%
\providecommand \bibitemNoStop [0]{.\EOS\space}%
\providecommand \EOS [0]{\spacefactor3000\relax}%
\providecommand \BibitemShut  [1]{\csname bibitem#1\endcsname}%
\let\auto@bib@innerbib\@empty
%</preamble>
\bibitem [{\citenamefont {Planck}(1914)}]{planck_theory_1914}%
  \BibitemOpen
  \bibfield  {author} {\bibinfo {author} {\bibfnamefont {M.}~\bibnamefont
  {Planck}},\ }\href {https://www.gutenberg.org/files/40030/40030-pdf.pdf}
  {\emph {\bibinfo {title} {The {Theory} of {Heat} {Radiation}}}}\ (\bibinfo
  {publisher} {P Blakiston's Son \& Co.},\ \bibinfo {address} {Philadelphia},\
  \bibinfo {year} {1914})\BibitemShut {NoStop}%
\bibitem [{\citenamefont {Nichols}\ and\ \citenamefont
  {Hull}(1901)}]{nichols_preliminary_1901}%
  \BibitemOpen
  \bibfield  {author} {\bibinfo {author} {\bibfnamefont {E.~F.}\ \bibnamefont
  {Nichols}}\ and\ \bibinfo {author} {\bibfnamefont {G.~F.}\ \bibnamefont
  {Hull}},\ }\href@noop {} {\bibfield  {journal} {\bibinfo  {journal} {Physical
  Review (Series I)}\ }\textbf {\bibinfo {volume} {13}},\ \bibinfo {pages}
  {307} (\bibinfo {year} {1901})}\BibitemShut {NoStop}%
\bibitem [{\citenamefont {Olmon}\ \emph {et~al.}(2012)\citenamefont {Olmon},
  \citenamefont {Slovick}, \citenamefont {Johnson}, \citenamefont {Shelton},
  \citenamefont {Oh}, \citenamefont {Boreman},\ and\ \citenamefont
  {Raschke}}]{olmon_optical_2012}%
  \BibitemOpen
  \bibfield  {author} {\bibinfo {author} {\bibfnamefont {R.~L.}\ \bibnamefont
  {Olmon}}, \bibinfo {author} {\bibfnamefont {B.}~\bibnamefont {Slovick}},
  \bibinfo {author} {\bibfnamefont {T.~W.}\ \bibnamefont {Johnson}}, \bibinfo
  {author} {\bibfnamefont {D.}~\bibnamefont {Shelton}}, \bibinfo {author}
  {\bibfnamefont {S.-H.}\ \bibnamefont {Oh}}, \bibinfo {author} {\bibfnamefont
  {G.~D.}\ \bibnamefont {Boreman}}, \ and\ \bibinfo {author} {\bibfnamefont
  {M.~B.}\ \bibnamefont {Raschke}},\ }\href {\doibase
  10.1103/PhysRevB.86.235147} {\bibfield  {journal} {\bibinfo  {journal}
  {Physical Review B}\ }\textbf {\bibinfo {volume} {86}},\ \bibinfo {pages}
  {235147} (\bibinfo {year} {2012})}\BibitemShut {NoStop}%
\bibitem [{\citenamefont {Wegener}(2005)}]{wegener_extreme_2005}%
  \BibitemOpen
  \bibfield  {author} {\bibinfo {author} {\bibfnamefont {M.}~\bibnamefont
  {Wegener}},\ }\href@noop {} {\emph {\bibinfo {title} {Extreme {Nonlinear}
  {Optics}}}}\ (\bibinfo  {publisher} {Springer},\ \bibinfo {address}
  {Berlin},\ \bibinfo {year} {2005})\BibitemShut {NoStop}%
\bibitem [{\citenamefont {Loudon}\ \emph {et~al.}(2005)\citenamefont {Loudon},
  \citenamefont {Barnett},\ and\ \citenamefont
  {Baxter}}]{loudon_radiation_2005}%
  \BibitemOpen
  \bibfield  {author} {\bibinfo {author} {\bibfnamefont {R.}~\bibnamefont
  {Loudon}}, \bibinfo {author} {\bibfnamefont {S.~M.}\ \bibnamefont {Barnett}},
  \ and\ \bibinfo {author} {\bibfnamefont {C.}~\bibnamefont {Baxter}},\ }\href
  {\doibase 10.1103/PhysRevA.71.063802} {\bibfield  {journal} {\bibinfo
  {journal} {Physical Review A}\ }\textbf {\bibinfo {volume} {71}},\ \bibinfo
  {pages} {063802} (\bibinfo {year} {2005})}\BibitemShut {NoStop}%
\bibitem [{\citenamefont {Goff}\ and\ \citenamefont
  {Schaich}(1997)}]{goff_hydrodynamic_1997}%
  \BibitemOpen
  \bibfield  {author} {\bibinfo {author} {\bibfnamefont {J.~E.}\ \bibnamefont
  {Goff}}\ and\ \bibinfo {author} {\bibfnamefont {W.~L.}\ \bibnamefont
  {Schaich}},\ }\href
  {https://journals.aps.org/prb/abstract/10.1103/PhysRevB.56.15421} {\bibfield
  {journal} {\bibinfo  {journal} {Physical Review B}\ }\textbf {\bibinfo
  {volume} {56}},\ \bibinfo {pages} {15421} (\bibinfo {year}
  {1997})}\BibitemShut {NoStop}%
\bibitem [{\citenamefont {Goff}\ and\ \citenamefont
  {Schaich}(2000)}]{goff_theory_2000}%
  \BibitemOpen
  \bibfield  {author} {\bibinfo {author} {\bibfnamefont {J.~E.}\ \bibnamefont
  {Goff}}\ and\ \bibinfo {author} {\bibfnamefont {W.~L.}\ \bibnamefont
  {Schaich}},\ }\href@noop {} {\bibfield  {journal} {\bibinfo  {journal}
  {Physical Review B}\ }\textbf {\bibinfo {volume} {61}},\ \bibinfo {pages}
  {10471} (\bibinfo {year} {2000})}\BibitemShut {NoStop}%
\bibitem [{\citenamefont {Gurevich}\ \emph {et~al.}(1992)\citenamefont
  {Gurevich}, \citenamefont {Laiho},\ and\ \citenamefont
  {Lashkul}}]{gurevich_photomagnetism_1992}%
  \BibitemOpen
  \bibfield  {author} {\bibinfo {author} {\bibfnamefont {V.~L.}\ \bibnamefont
  {Gurevich}}, \bibinfo {author} {\bibfnamefont {R.}~\bibnamefont {Laiho}}, \
  and\ \bibinfo {author} {\bibfnamefont {A.~V.}\ \bibnamefont {Lashkul}},\
  }\href {https://journals.aps.org/prl/abstract/10.1103/PhysRevLett.69.180}
  {\bibfield  {journal} {\bibinfo  {journal} {Physical Review Letters}\
  }\textbf {\bibinfo {volume} {69}},\ \bibinfo {pages} {180} (\bibinfo {year}
  {1992})}\BibitemShut {NoStop}%
\bibitem [{\citenamefont {Durach}\ \emph {et~al.}(2009)\citenamefont {Durach},
  \citenamefont {Rusina},\ and\ \citenamefont {Stockman}}]{durach_giant_2009}%
  \BibitemOpen
  \bibfield  {author} {\bibinfo {author} {\bibfnamefont {M.}~\bibnamefont
  {Durach}}, \bibinfo {author} {\bibfnamefont {A.}~\bibnamefont {Rusina}}, \
  and\ \bibinfo {author} {\bibfnamefont {M.~I.}\ \bibnamefont {Stockman}},\
  }\href {\doibase 10.1103/PhysRevLett.103.186801} {\bibfield  {journal}
  {\bibinfo  {journal} {Physical Review Letters}\ }\textbf {\bibinfo {volume}
  {103}},\ \bibinfo {pages} {186801} (\bibinfo {year} {2009})}\BibitemShut
  {NoStop}%
\bibitem [{\citenamefont {Bai}(2015)}]{bai_manipulating_2015}%
  \BibitemOpen
  \bibfield  {author} {\bibinfo {author} {\bibfnamefont {Q.}~\bibnamefont
  {Bai}},\ }\href {\doibase 10.1364/OE.23.005348} {\bibfield  {journal}
  {\bibinfo  {journal} {Optics Express}\ }\textbf {\bibinfo {volume} {23}},\
  \bibinfo {pages} {5348} (\bibinfo {year} {2015})}\BibitemShut {NoStop}%
\bibitem [{\citenamefont {Durach}\ and\ \citenamefont
  {Noginova}(2017)}]{durach_spin_2017}%
  \BibitemOpen
  \bibfield  {author} {\bibinfo {author} {\bibfnamefont {M.}~\bibnamefont
  {Durach}}\ and\ \bibinfo {author} {\bibfnamefont {N.}~\bibnamefont
  {Noginova}},\ }\href {\doibase 10.1103/PhysRevB.96.195411} {\bibfield
  {journal} {\bibinfo  {journal} {Physical Review B}\ }\textbf {\bibinfo
  {volume} {96}},\ \bibinfo {pages} {195411} (\bibinfo {year}
  {2017})}\BibitemShut {NoStop}%
\bibitem [{\citenamefont {Lupi}(2014)}]{lupi_electrical_2014}%
  \BibitemOpen
  \bibfield  {author} {\bibinfo {author} {\bibfnamefont {A.}~\bibnamefont
  {Lupi}},\ }\emph {\bibinfo {title} {Electrical {Detection} of {Surface}
  {Plasmon} {Polaritons} via the {Plasmon} {Drag} {Effect}}},\ \href@noop {}
  {Ph.D. thesis},\ \bibinfo  {school} {Imperial College London} (\bibinfo
  {year} {2014})\BibitemShut {NoStop}%
\bibitem [{\citenamefont {Vengurlekar}\ and\ \citenamefont
  {Ishihara}(2005)}]{vengurlekar_surface_2005}%
  \BibitemOpen
  \bibfield  {author} {\bibinfo {author} {\bibfnamefont {A.~S.}\ \bibnamefont
  {Vengurlekar}}\ and\ \bibinfo {author} {\bibfnamefont {T.}~\bibnamefont
  {Ishihara}},\ }\href {\doibase 10.1063/1.2037851} {\bibfield  {journal}
  {\bibinfo  {journal} {Applied Physics Letters}\ }\textbf {\bibinfo {volume}
  {87}},\ \bibinfo {pages} {091118} (\bibinfo {year} {2005})}\BibitemShut
  {NoStop}%
\bibitem [{\citenamefont {Noginova}\ \emph {et~al.}(2011)\citenamefont
  {Noginova}, \citenamefont {Yakim}, \citenamefont {Soimo}, \citenamefont
  {Gu},\ and\ \citenamefont {Noginov}}]{noginova_light--current_2011}%
  \BibitemOpen
  \bibfield  {author} {\bibinfo {author} {\bibfnamefont {N.}~\bibnamefont
  {Noginova}}, \bibinfo {author} {\bibfnamefont {A.~V.}\ \bibnamefont {Yakim}},
  \bibinfo {author} {\bibfnamefont {J.}~\bibnamefont {Soimo}}, \bibinfo
  {author} {\bibfnamefont {L.}~\bibnamefont {Gu}}, \ and\ \bibinfo {author}
  {\bibfnamefont {M.~A.}\ \bibnamefont {Noginov}},\ }\href {\doibase
  10.1103/PhysRevB.84.035447} {\bibfield  {journal} {\bibinfo  {journal}
  {Physical Review B}\ }\textbf {\bibinfo {volume} {84}},\ \bibinfo {pages}
  {035447} (\bibinfo {year} {2011})}\BibitemShut {NoStop}%
\bibitem [{\citenamefont {Durach}\ and\ \citenamefont
  {Noginova}(2016)}]{durach_nature_2016}%
  \BibitemOpen
  \bibfield  {author} {\bibinfo {author} {\bibfnamefont {M.}~\bibnamefont
  {Durach}}\ and\ \bibinfo {author} {\bibfnamefont {N.}~\bibnamefont
  {Noginova}},\ }\href {\doibase 10.1103/PhysRevB.93.161406} {\bibfield
  {journal} {\bibinfo  {journal} {Physical Review B}\ }\textbf {\bibinfo
  {volume} {93}},\ \bibinfo {pages} {161406} (\bibinfo {year}
  {2016})}\BibitemShut {NoStop}%
\bibitem [{\citenamefont {Noginova}\ \emph {et~al.}(2013)\citenamefont
  {Noginova}, \citenamefont {Rono}, \citenamefont {Bezares},\ and\
  \citenamefont {Caldwell}}]{noginova_plasmon_2013}%
  \BibitemOpen
  \bibfield  {author} {\bibinfo {author} {\bibfnamefont {N.}~\bibnamefont
  {Noginova}}, \bibinfo {author} {\bibfnamefont {V.}~\bibnamefont {Rono}},
  \bibinfo {author} {\bibfnamefont {F.~J.}\ \bibnamefont {Bezares}}, \ and\
  \bibinfo {author} {\bibfnamefont {J.~D.}\ \bibnamefont {Caldwell}},\ }\href
  {\doibase 10.1088/1367-2630/15/11/113061} {\bibfield  {journal} {\bibinfo
  {journal} {New Journal of Physics}\ }\textbf {\bibinfo {volume} {15}},\
  \bibinfo {pages} {113061} (\bibinfo {year} {2013})}\BibitemShut {NoStop}%
\bibitem [{\citenamefont {Hatano}\ \emph {et~al.}(2009)\citenamefont {Hatano},
  \citenamefont {Ishihara}, \citenamefont {Tikhodeev},\ and\ \citenamefont
  {Gippius}}]{hatano_transverse_2009}%
  \BibitemOpen
  \bibfield  {author} {\bibinfo {author} {\bibfnamefont {T.}~\bibnamefont
  {Hatano}}, \bibinfo {author} {\bibfnamefont {T.}~\bibnamefont {Ishihara}},
  \bibinfo {author} {\bibfnamefont {S.~G.}\ \bibnamefont {Tikhodeev}}, \ and\
  \bibinfo {author} {\bibfnamefont {N.~A.}\ \bibnamefont {Gippius}},\ }\href
  {\doibase 10.1103/PhysRevLett.103.103906} {\bibfield  {journal} {\bibinfo
  {journal} {Physical Review Letters}\ }\textbf {\bibinfo {volume} {103}},\
  \bibinfo {pages} {103906} (\bibinfo {year} {2009})}\BibitemShut {NoStop}%
\bibitem [{\citenamefont {Akbari}\ \emph {et~al.}(2015)\citenamefont {Akbari},
  \citenamefont {Onoda},\ and\ \citenamefont
  {Ishihara}}]{akbari_photo-induced_2015}%
  \BibitemOpen
  \bibfield  {author} {\bibinfo {author} {\bibfnamefont {M.}~\bibnamefont
  {Akbari}}, \bibinfo {author} {\bibfnamefont {M.}~\bibnamefont {Onoda}}, \
  and\ \bibinfo {author} {\bibfnamefont {T.}~\bibnamefont {Ishihara}},\ }\href
  {\doibase 10.1364/OE.23.000823} {\bibfield  {journal} {\bibinfo  {journal}
  {Optics Express}\ }\textbf {\bibinfo {volume} {23}},\ \bibinfo {pages} {823}
  (\bibinfo {year} {2015})}\BibitemShut {NoStop}%
\bibitem [{\citenamefont {Hatano}\ \emph {et~al.}(2008)\citenamefont {Hatano},
  \citenamefont {Nishikawa}, \citenamefont {Iwanaga},\ and\ \citenamefont
  {Ishihara}}]{hatano_optical_2008}%
  \BibitemOpen
  \bibfield  {author} {\bibinfo {author} {\bibfnamefont {T.}~\bibnamefont
  {Hatano}}, \bibinfo {author} {\bibfnamefont {B.}~\bibnamefont {Nishikawa}},
  \bibinfo {author} {\bibfnamefont {M.}~\bibnamefont {Iwanaga}}, \ and\
  \bibinfo {author} {\bibfnamefont {T.}~\bibnamefont {Ishihara}},\ }\href
  {\doibase 10.1364/OE.16.008236} {\bibfield  {journal} {\bibinfo  {journal}
  {Optics Express}\ }\textbf {\bibinfo {volume} {16}},\ \bibinfo {pages} {8236}
  (\bibinfo {year} {2008})}\BibitemShut {NoStop}%
\bibitem [{\citenamefont {Kurosawa}\ \emph {et~al.}(2012)\citenamefont
  {Kurosawa}, \citenamefont {Ishihara}, \citenamefont {Ikeda}, \citenamefont
  {Tsuya}, \citenamefont {Ochiai},\ and\ \citenamefont
  {Sugimoto}}]{kurosawa_optical_2012}%
  \BibitemOpen
  \bibfield  {author} {\bibinfo {author} {\bibfnamefont {H.}~\bibnamefont
  {Kurosawa}}, \bibinfo {author} {\bibfnamefont {T.}~\bibnamefont {Ishihara}},
  \bibinfo {author} {\bibfnamefont {N.}~\bibnamefont {Ikeda}}, \bibinfo
  {author} {\bibfnamefont {D.}~\bibnamefont {Tsuya}}, \bibinfo {author}
  {\bibfnamefont {M.}~\bibnamefont {Ochiai}}, \ and\ \bibinfo {author}
  {\bibfnamefont {Y.}~\bibnamefont {Sugimoto}},\ }\href@noop {} {\bibfield
  {journal} {\bibinfo  {journal} {Optics letters}\ }\textbf {\bibinfo {volume}
  {37}},\ \bibinfo {pages} {2793} (\bibinfo {year} {2012})}\BibitemShut
  {NoStop}%
\bibitem [{\citenamefont {Kurosawa}\ and\ \citenamefont
  {Ishihara}(2012)}]{kurosawa_surface_2012}%
  \BibitemOpen
  \bibfield  {author} {\bibinfo {author} {\bibfnamefont {H.}~\bibnamefont
  {Kurosawa}}\ and\ \bibinfo {author} {\bibfnamefont {T.}~\bibnamefont
  {Ishihara}},\ }\href {\doibase 10.1364/OE.20.001561} {\bibfield  {journal}
  {\bibinfo  {journal} {Optics Express}\ }\textbf {\bibinfo {volume} {20}},\
  \bibinfo {pages} {1561} (\bibinfo {year} {2012})}\BibitemShut {NoStop}%
\bibitem [{\citenamefont {Noginova}\ \emph {et~al.}(2016)\citenamefont
  {Noginova}, \citenamefont {LePain}, \citenamefont {Rono}, \citenamefont
  {Mashhadi}, \citenamefont {Hussain},\ and\ \citenamefont
  {Durach}}]{noginova_plasmonic_2016}%
  \BibitemOpen
  \bibfield  {author} {\bibinfo {author} {\bibfnamefont {N.}~\bibnamefont
  {Noginova}}, \bibinfo {author} {\bibfnamefont {M.}~\bibnamefont {LePain}},
  \bibinfo {author} {\bibfnamefont {V.}~\bibnamefont {Rono}}, \bibinfo {author}
  {\bibfnamefont {S.}~\bibnamefont {Mashhadi}}, \bibinfo {author}
  {\bibfnamefont {R.}~\bibnamefont {Hussain}}, \ and\ \bibinfo {author}
  {\bibfnamefont {M.}~\bibnamefont {Durach}},\ }\href {\doibase
  10.1088/1367-2630/18/9/093036} {\bibfield  {journal} {\bibinfo  {journal}
  {New Journal of Physics}\ }\textbf {\bibinfo {volume} {18}},\ \bibinfo
  {pages} {093036} (\bibinfo {year} {2016})}\BibitemShut {NoStop}%
\bibitem [{\citenamefont {Proscia}\ \emph {et~al.}(2016)\citenamefont
  {Proscia}, \citenamefont {Moocarme}, \citenamefont {Chang}, \citenamefont
  {Kretzschmar}, \citenamefont {Menon},\ and\ \citenamefont
  {Vuong}}]{proscia_control_2016}%
  \BibitemOpen
  \bibfield  {author} {\bibinfo {author} {\bibfnamefont {N.~V.}\ \bibnamefont
  {Proscia}}, \bibinfo {author} {\bibfnamefont {M.}~\bibnamefont {Moocarme}},
  \bibinfo {author} {\bibfnamefont {R.}~\bibnamefont {Chang}}, \bibinfo
  {author} {\bibfnamefont {I.}~\bibnamefont {Kretzschmar}}, \bibinfo {author}
  {\bibfnamefont {V.~M.}\ \bibnamefont {Menon}}, \ and\ \bibinfo {author}
  {\bibfnamefont {L.~T.}\ \bibnamefont {Vuong}},\ }\href {\doibase
  10.1364/OE.24.010402} {\bibfield  {journal} {\bibinfo  {journal} {Optics
  Express}\ }\textbf {\bibinfo {volume} {24}},\ \bibinfo {pages} {10402}
  (\bibinfo {year} {2016})}\BibitemShut {NoStop}%
\bibitem [{\citenamefont {Akbari}\ and\ \citenamefont
  {Ishihara}(2017)}]{akbari_polarization_2017}%
  \BibitemOpen
  \bibfield  {author} {\bibinfo {author} {\bibfnamefont {M.}~\bibnamefont
  {Akbari}}\ and\ \bibinfo {author} {\bibfnamefont {T.}~\bibnamefont
  {Ishihara}},\ }\href {\doibase 10.1364/OE.25.002143} {\bibfield  {journal}
  {\bibinfo  {journal} {Optics Express}\ }\textbf {\bibinfo {volume} {25}},\
  \bibinfo {pages} {2143} (\bibinfo {year} {2017})}\BibitemShut {NoStop}%
\bibitem [{\citenamefont {Akbari}\ \emph {et~al.}(2018)\citenamefont {Akbari},
  \citenamefont {Gao},\ and\ \citenamefont {Yang}}]{akbari_generation_2018}%
  \BibitemOpen
  \bibfield  {author} {\bibinfo {author} {\bibfnamefont {M.}~\bibnamefont
  {Akbari}}, \bibinfo {author} {\bibfnamefont {J.}~\bibnamefont {Gao}}, \ and\
  \bibinfo {author} {\bibfnamefont {X.}~\bibnamefont {Yang}},\ }\href {\doibase
  10.1364/OE.26.021194} {\bibfield  {journal} {\bibinfo  {journal} {Optics
  Express}\ }\textbf {\bibinfo {volume} {26}},\ \bibinfo {pages} {21194}
  (\bibinfo {year} {2018})}\BibitemShut {NoStop}%
\bibitem [{\citenamefont {Ashcroft}\ and\ \citenamefont
  {Mermin}(1976)}]{ashcroft_solid_1976}%
  \BibitemOpen
  \bibfield  {author} {\bibinfo {author} {\bibfnamefont {N.~W.}\ \bibnamefont
  {Ashcroft}}\ and\ \bibinfo {author} {\bibfnamefont {N.~D.}\ \bibnamefont
  {Mermin}},\ }\href@noop {} {\emph {\bibinfo {title} {Solid {State}
  {Physics}}}}\ (\bibinfo  {publisher} {Brooks/Cole},\ \bibinfo {address}
  {Belmont, CA},\ \bibinfo {year} {1976})\BibitemShut {NoStop}%
\bibitem [{\citenamefont {Lompre}\ \emph {et~al.}(1978)\citenamefont {Lompre},
  \citenamefont {Mainfray}, \citenamefont {Manus}, \citenamefont {Thebault},
  \citenamefont {Farkas},\ and\ \citenamefont {Horvath}}]{lompre_new_1978}%
  \BibitemOpen
  \bibfield  {author} {\bibinfo {author} {\bibfnamefont {L.~A.}\ \bibnamefont
  {Lompre}}, \bibinfo {author} {\bibfnamefont {G.}~\bibnamefont {Mainfray}},
  \bibinfo {author} {\bibfnamefont {C.}~\bibnamefont {Manus}}, \bibinfo
  {author} {\bibfnamefont {J.}~\bibnamefont {Thebault}}, \bibinfo {author}
  {\bibfnamefont {G.}~\bibnamefont {Farkas}}, \ and\ \bibinfo {author}
  {\bibfnamefont {Z.}~\bibnamefont {Horvath}},\ }\href {\doibase
  10.1063/1.90306} {\bibfield  {journal} {\bibinfo  {journal} {Applied Physics
  Letters}\ }\textbf {\bibinfo {volume} {33}},\ \bibinfo {pages} {124}
  (\bibinfo {year} {1978})}\BibitemShut {NoStop}%
\bibitem [{noa()}]{noauthor_see_nodate}%
  \BibitemOpen
  \href@noop {} {\enquote {\bibinfo {title} {See Supplemental Material at … for description of an electrical sign-verification experiment.},}\ }\BibitemShut {NoStop}%
\bibitem [{\citenamefont {Kats}\ and\ \citenamefont
  {Capasso}(2016)}]{kats_optical_2016}%
  \BibitemOpen
  \bibfield  {author} {\bibinfo {author} {\bibfnamefont {M.~A.}\ \bibnamefont
  {Kats}}\ and\ \bibinfo {author} {\bibfnamefont {F.}~\bibnamefont {Capasso}},\
  }\href {\doibase 10.1002/lpor.201600098} {\bibfield  {journal} {\bibinfo
  {journal} {Laser \& Photonics Reviews}\ }\textbf {\bibinfo {volume} {10}},\
  \bibinfo {pages} {735} (\bibinfo {year} {2016})}\BibitemShut {NoStop}%
\bibitem [{\citenamefont {Raether}(1988)}]{raether_surface_1988}%
  \BibitemOpen
  \bibfield  {author} {\bibinfo {author} {\bibfnamefont {H.}~\bibnamefont
  {Raether}},\ }\href {//www.springer.com/us/book/9783662151242} {\emph
  {\bibinfo {title} {Surface {Plasmons} on {Smooth} and {Rough} {Surfaces} and
  on {Gratings}}}}\ (\bibinfo  {publisher} {Springer-Verlag},\ \bibinfo
  {address} {Berlin Heidelberg},\ \bibinfo {year} {1988})\BibitemShut {NoStop}%
\bibitem [{\citenamefont {Kadlec}\ \emph {et~al.}(2005)\citenamefont {Kadlec},
  \citenamefont {Kuzel},\ and\ \citenamefont {Coutaz}}]{kadlec_study_2005}%
  \BibitemOpen
  \bibfield  {author} {\bibinfo {author} {\bibfnamefont {F.}~\bibnamefont
  {Kadlec}}, \bibinfo {author} {\bibfnamefont {P.}~\bibnamefont {Kuzel}}, \
  and\ \bibinfo {author} {\bibfnamefont {J.-L.}\ \bibnamefont {Coutaz}},\
  }\href@noop {} {\bibfield  {journal} {\bibinfo  {journal} {Optics letters}\
  }\textbf {\bibinfo {volume} {30}},\ \bibinfo {pages} {1402} (\bibinfo {year}
  {2005})}\BibitemShut {NoStop}%
\bibitem [{\citenamefont {Hubner}\ \emph {et~al.}(1994)\citenamefont {Hubner},
  \citenamefont {Bennemann},\ and\ \citenamefont
  {Bohmer}}]{hubner_theory_1994}%
  \BibitemOpen
  \bibfield  {author} {\bibinfo {author} {\bibfnamefont {W.}~\bibnamefont
  {Hubner}}, \bibinfo {author} {\bibfnamefont {K.~H.}\ \bibnamefont
  {Bennemann}}, \ and\ \bibinfo {author} {\bibfnamefont {K.}~\bibnamefont
  {Bohmer}},\ }\href@noop {} {\bibfield  {journal} {\bibinfo  {journal}
  {Physical Review B}\ }\textbf {\bibinfo {volume} {50}},\ \bibinfo {pages}
  {17597} (\bibinfo {year} {1994})}\BibitemShut {NoStop}%
\bibitem [{\citenamefont {Gurevich}\ and\ \citenamefont
  {Laiho}(1993)}]{gurevich_photomagnetism_1993}%
  \BibitemOpen
  \bibfield  {author} {\bibinfo {author} {\bibfnamefont {V.~L.}\ \bibnamefont
  {Gurevich}}\ and\ \bibinfo {author} {\bibfnamefont {R.}~\bibnamefont
  {Laiho}},\ }\href
  {https://journals.aps.org/prb/abstract/10.1103/PhysRevB.48.8307} {\bibfield
  {journal} {\bibinfo  {journal} {Physical Review B}\ }\textbf {\bibinfo
  {volume} {48}},\ \bibinfo {pages} {8307} (\bibinfo {year}
  {1993})}\BibitemShut {NoStop}%
\bibitem [{\citenamefont {Gurevich}\ and\ \citenamefont
  {Laiho}(2000)}]{gurevich_photomagnetism_2000}%
  \BibitemOpen
  \bibfield  {author} {\bibinfo {author} {\bibfnamefont {V.~L.}\ \bibnamefont
  {Gurevich}}\ and\ \bibinfo {author} {\bibfnamefont {R.}~\bibnamefont
  {Laiho}},\ }\href {http://link.springer.com/article/10.1134/1.1318868}
  {\bibfield  {journal} {\bibinfo  {journal} {Physics of the Solid State}\
  }\textbf {\bibinfo {volume} {42}},\ \bibinfo {pages} {1807} (\bibinfo {year}
  {2000})}\BibitemShut {NoStop}%
\bibitem [{\citenamefont {Mikheev}\ \emph {et~al.}(2018)\citenamefont
  {Mikheev}, \citenamefont {Saushin}, \citenamefont {Styapshin},\ and\
  \citenamefont {Svirko}}]{mikheev_interplay_2018}%
  \BibitemOpen
  \bibfield  {author} {\bibinfo {author} {\bibfnamefont {G.~M.}\ \bibnamefont
  {Mikheev}}, \bibinfo {author} {\bibfnamefont {A.~S.}\ \bibnamefont
  {Saushin}}, \bibinfo {author} {\bibfnamefont {V.~M.}\ \bibnamefont
  {Styapshin}}, \ and\ \bibinfo {author} {\bibfnamefont {Y.~P.}\ \bibnamefont
  {Svirko}},\ }\href {\doibase 10.1038/s41598-018-26923-2} {\bibfield
  {journal} {\bibinfo  {journal} {Scientific Reports}\ }\textbf {\bibinfo
  {volume} {8}},\ \bibinfo {pages} {8644} (\bibinfo {year} {2018})}\BibitemShut
  {NoStop}%
\bibitem [{\citenamefont {Gibson}\ \emph {et~al.}(1980)\citenamefont {Gibson},
  \citenamefont {Kimmitt}, \citenamefont {Koohian}, \citenamefont {Evans},\
  and\ \citenamefont {Levy}}]{gibson_study_1980}%
  \BibitemOpen
  \bibfield  {author} {\bibinfo {author} {\bibfnamefont {A.~F.}\ \bibnamefont
  {Gibson}}, \bibinfo {author} {\bibfnamefont {M.~F.}\ \bibnamefont {Kimmitt}},
  \bibinfo {author} {\bibfnamefont {A.~O.}\ \bibnamefont {Koohian}}, \bibinfo
  {author} {\bibfnamefont {D.~E.}\ \bibnamefont {Evans}}, \ and\ \bibinfo
  {author} {\bibfnamefont {G.~F.~D.}\ \bibnamefont {Levy}},\ }\href
  {http://www.jstor.org/stable/2990241} {\bibfield  {journal} {\bibinfo
  {journal} {Proceedings of the Royal Society of London. Series A, Mathematical
  and Physical Sciences}\ }\textbf {\bibinfo {volume} {370}},\ \bibinfo {pages}
  {303} (\bibinfo {year} {1980})}\BibitemShut {NoStop}%
\bibitem [{\citenamefont {Wieck}\ \emph {et~al.}(1990)\citenamefont {Wieck},
  \citenamefont {Sigg},\ and\ \citenamefont {Ploog}}]{wieck_observation_1990}%
  \BibitemOpen
  \bibfield  {author} {\bibinfo {author} {\bibfnamefont {A.~D.}\ \bibnamefont
  {Wieck}}, \bibinfo {author} {\bibfnamefont {H.}~\bibnamefont {Sigg}}, \ and\
  \bibinfo {author} {\bibfnamefont {K.}~\bibnamefont {Ploog}},\ }\href
  {\doibase 10.1103/PhysRevLett.64.463} {\bibfield  {journal} {\bibinfo
  {journal} {Physical Review Letters}\ }\textbf {\bibinfo {volume} {64}},\
  \bibinfo {pages} {463} (\bibinfo {year} {1990})}\BibitemShut {NoStop}%
\bibitem [{\citenamefont {Gibson}\ and\ \citenamefont
  {Montasser}(1975)}]{gibson_theoretical_1975}%
  \BibitemOpen
  \bibfield  {author} {\bibinfo {author} {\bibfnamefont {A.~F.}\ \bibnamefont
  {Gibson}}\ and\ \bibinfo {author} {\bibfnamefont {S.}~\bibnamefont
  {Montasser}},\ }\href
  {http://iopscience.iop.org/article/10.1088/0022-3719/8/19/015/meta}
  {\bibfield  {journal} {\bibinfo  {journal} {Journal of Physics C: Solid State
  Physics}\ }\textbf {\bibinfo {volume} {8}},\ \bibinfo {pages} {3147}
  (\bibinfo {year} {1975})}\BibitemShut {NoStop}%
\bibitem [{\citenamefont {Luryi}(1987)}]{luryi_photon-drag_1987}%
  \BibitemOpen
  \bibfield  {author} {\bibinfo {author} {\bibfnamefont {S.}~\bibnamefont
  {Luryi}},\ }\href {\doibase 10.1103/PhysRevLett.58.2263} {\bibfield
  {journal} {\bibinfo  {journal} {Physical Review Letters}\ }\textbf {\bibinfo
  {volume} {58}},\ \bibinfo {pages} {2263} (\bibinfo {year}
  {1987})}\BibitemShut {NoStop}%
\bibitem [{\citenamefont {Grinberg}\ and\ \citenamefont
  {Luryi}(1988)}]{grinberg_theory_1988}%
  \BibitemOpen
  \bibfield  {author} {\bibinfo {author} {\bibfnamefont {A.~A.}\ \bibnamefont
  {Grinberg}}\ and\ \bibinfo {author} {\bibfnamefont {S.}~\bibnamefont
  {Luryi}},\ }\href
  {http://journals.aps.org/prb/abstract/10.1103/PhysRevB.38.87} {\bibfield
  {journal} {\bibinfo  {journal} {Physical Review B}\ }\textbf {\bibinfo
  {volume} {38}},\ \bibinfo {pages} {87} (\bibinfo {year} {1988})}\BibitemShut
  {NoStop}%
\bibitem [{\citenamefont {Shalaev}\ \emph {et~al.}(1996)\citenamefont
  {Shalaev}, \citenamefont {Douketis}, \citenamefont {Stuckless},\ and\
  \citenamefont {Moskovits}}]{shalaev_light-induced_1996}%
  \BibitemOpen
  \bibfield  {author} {\bibinfo {author} {\bibfnamefont {V.~M.}\ \bibnamefont
  {Shalaev}}, \bibinfo {author} {\bibfnamefont {C.}~\bibnamefont {Douketis}},
  \bibinfo {author} {\bibfnamefont {J.~T.}\ \bibnamefont {Stuckless}}, \ and\
  \bibinfo {author} {\bibfnamefont {M.}~\bibnamefont {Moskovits}},\ }\href@noop
  {} {\bibfield  {journal} {\bibinfo  {journal} {Physical Review B}\ }\textbf
  {\bibinfo {volume} {53}},\ \bibinfo {pages} {11388} (\bibinfo {year}
  {1996})}\BibitemShut {NoStop}%
\bibitem [{\citenamefont {Barnett}(2010)}]{barnett_resolution_2010}%
  \BibitemOpen
  \bibfield  {author} {\bibinfo {author} {\bibfnamefont {S.~M.}\ \bibnamefont
  {Barnett}},\ }\href {\doibase 10.1103/PhysRevLett.104.070401} {\bibfield
  {journal} {\bibinfo  {journal} {Physical Review Letters}\ }\textbf {\bibinfo
  {volume} {104}},\ \bibinfo {pages} {070401} (\bibinfo {year}
  {2010})}\BibitemShut {NoStop}%
\bibitem [{\citenamefont {Partanen}\ \emph {et~al.}(2017)\citenamefont
  {Partanen}, \citenamefont {Hayrynen}, \citenamefont {Oksanen},\ and\
  \citenamefont {Tulkki}}]{partanen_photon_2017}%
  \BibitemOpen
  \bibfield  {author} {\bibinfo {author} {\bibfnamefont {M.}~\bibnamefont
  {Partanen}}, \bibinfo {author} {\bibfnamefont {T.}~\bibnamefont {Hayrynen}},
  \bibinfo {author} {\bibfnamefont {J.}~\bibnamefont {Oksanen}}, \ and\
  \bibinfo {author} {\bibfnamefont {J.}~\bibnamefont {Tulkki}},\ }\href
  {\doibase 10.1103/PhysRevA.95.063850} {\bibfield  {journal} {\bibinfo
  {journal} {Physical Review A}\ }\textbf {\bibinfo {volume} {95}},\ \bibinfo
  {pages} {063850} (\bibinfo {year} {2017})}\BibitemShut {NoStop}%
\bibitem [{\citenamefont {Wells}\ and\ \citenamefont
  {Fort}(1972)}]{wells_adsorption_1972}%
  \BibitemOpen
  \bibfield  {author} {\bibinfo {author} {\bibfnamefont {R.~L.}\ \bibnamefont
  {Wells}}\ and\ \bibinfo {author} {\bibfnamefont {T.}~\bibnamefont {Fort}},\
  }\href {\doibase 10.1016/0039-6028(72)90182-3} {\bibfield  {journal}
  {\bibinfo  {journal} {Surface Science}\ }\textbf {\bibinfo {volume} {32}},\
  \bibinfo {pages} {554} (\bibinfo {year} {1972})}\BibitemShut {NoStop}%
\bibitem [{\citenamefont {Demirdjian}\ \emph {et~al.}(2018)\citenamefont
  {Demirdjian}, \citenamefont {Bedu}, \citenamefont {Ranguis}, \citenamefont
  {Ozerov},\ and\ \citenamefont {Henry}}]{demirdjian_water_2018}%
  \BibitemOpen
  \bibfield  {author} {\bibinfo {author} {\bibfnamefont {B.}~\bibnamefont
  {Demirdjian}}, \bibinfo {author} {\bibfnamefont {F.}~\bibnamefont {Bedu}},
  \bibinfo {author} {\bibfnamefont {A.}~\bibnamefont {Ranguis}}, \bibinfo
  {author} {\bibfnamefont {I.}~\bibnamefont {Ozerov}}, \ and\ \bibinfo {author}
  {\bibfnamefont {C.~R.}\ \bibnamefont {Henry}},\ }\href {\doibase
  10.1021/acs.langmuir.8b00040} {\bibfield  {journal} {\bibinfo  {journal}
  {Langmuir}\ }\textbf {\bibinfo {volume} {34}},\ \bibinfo {pages} {5381}
  (\bibinfo {year} {2018})}\BibitemShut {NoStop}%
\bibitem [{\citenamefont {Brown}\ and\ \citenamefont
  {Matsuoka}(1969)}]{brown_effect_1969}%
  \BibitemOpen
  \bibfield  {author} {\bibinfo {author} {\bibfnamefont {F.}~\bibnamefont
  {Brown}}\ and\ \bibinfo {author} {\bibfnamefont {M.}~\bibnamefont
  {Matsuoka}},\ }\href {\doibase 10.1103/PhysRev.185.985} {\bibfield  {journal}
  {\bibinfo  {journal} {Physical Review}\ }\textbf {\bibinfo {volume} {185}},\
  \bibinfo {pages} {985} (\bibinfo {year} {1969})}\BibitemShut {NoStop}%
\bibitem [{\citenamefont {Shen}(1989)}]{shen_surface_1989}%
  \BibitemOpen
  \bibfield  {author} {\bibinfo {author} {\bibfnamefont {Y.~R.}\ \bibnamefont
  {Shen}},\ }\href {https://www.nature.com/articles/337519a0.pdf} {\bibfield
  {journal} {\bibinfo  {journal} {Nature}\ }\textbf {\bibinfo {volume} {337}},\
  \bibinfo {pages} {519} (\bibinfo {year} {1989})}\BibitemShut {NoStop}%
\bibitem [{\citenamefont {Mansuripur}\ \emph {et~al.}(2013)\citenamefont
  {Mansuripur}, \citenamefont {Zakharian},\ and\ \citenamefont
  {Wright}}]{mansuripur_electromagnetic-force_2013}%
  \BibitemOpen
  \bibfield  {author} {\bibinfo {author} {\bibfnamefont {M.}~\bibnamefont
  {Mansuripur}}, \bibinfo {author} {\bibfnamefont {A.~R.}\ \bibnamefont
  {Zakharian}}, \ and\ \bibinfo {author} {\bibfnamefont {E.~M.}\ \bibnamefont
  {Wright}},\ }\href {\doibase 10.1103/PhysRevA.88.023826} {\bibfield
  {journal} {\bibinfo  {journal} {Physical Review A}\ }\textbf {\bibinfo
  {volume} {88}},\ \bibinfo {pages} {023826} (\bibinfo {year}
  {2013})}\BibitemShut {NoStop}%
\bibitem [{\citenamefont {Zangwill}(2012)}]{zangwill_modern_2012}%
  \BibitemOpen
  \bibfield  {author} {\bibinfo {author} {\bibfnamefont {A.}~\bibnamefont
  {Zangwill}},\ }\href
  {http://www.cambridge.org/us/academic/subjects/physics/general-and-classical-physics/modern-electrodynamics}
  {\emph {\bibinfo {title} {Modern electrodynamics {\textbar} {General} and
  classical physics}}}\ (\bibinfo  {publisher} {Cambridge University Press},\
  \bibinfo {year} {2012})\BibitemShut {NoStop}%
\end{thebibliography}
\end{document}